\documentclass[11pt]{article}
\usepackage{graphicx}
\usepackage{amsmath}
\begin{document}

\begin{center} 
{\Large\bf   QCD Factorization of Quasi Generalized Quark Distributions  }
\par\vskip20pt
 J.P. Ma$^{1,2,3}$,  Z.Y. Pang$^{1,2}$ and G.P. Zhang$^{4}$    \\
{\small {\it
$^1$ CAS Key Laboratory of Theoretical Physics, Institute of Theoretical Physics, P.O. Box 2735, Chinese Academy of Sciences, Beijing 100190, China\\
$^2$ School of Physical Sciences, University of Chinese Academy of Sciences, Beijing 100049, China\\
$^3$ School of Physics and Center for High-Energy Physics, Peking University, Beijing 100871, China\\
$^4$ Department of Physics, Yunnan University, Kunming, Yunnan 650091, China}} \\
\end{center}

\vskip 1cm
\begin{abstract}
We study the factorization of quasi generalized quark distributions with  twist-2 generalized parton distributions. We use an approach which is different than that used in literature. Using the approach we 
derive the factorization relations of all quasi generalized quark distributions at one-loop.  The contributions from twist-2 generalized gluon distributions are included.  Our results apply not only to the quasi distributions of a spin-1/2 hadron but also to those of a hadron with any spin.

\end{abstract}      
\vskip 5mm

\vskip40pt

\noindent 
{\bf 1. Introduction}
	
\par

Parton distributions are important nonperturbative quantities in QCD studies. These distributions describe motions of partons
inside a hadron and hence contain information about inner structure of hadrons. The most familiar distributions are twist-2 Parton Distribution Functions(PDFs) which are used to predict differential cross sections of hadron collisions with large momentum transfers like those measured with LHC experiments.  
Therefore, it is important to know these distributions in detail. 
 
Since they contain nonperturbative effects, parton distributions can only be calculated with nonperturbative methods in theory or extracted from experimental data. With Lattice QCD one can not calculate
these distributions directly, because they depend explicitly on time. Only 
moments of these distributions as matrix elements of local operators can be calculated with Lattice QCD.  
Recently, a new method, or the so-called Large-Momentum Effective Field Theory(LaMET),  has been proposed to calculate twist-2 parton distributions directly in \cite{Ji,JiC}.
With this method one uses time-independent operators to define quasi parton distributions of a hadron which can be conveniently calculated with Lattice QCD.  In the limit of large hadron momentum one can show that these quasi parton distributions can be related to twist-2 parton distributions. The relation 
is perturbative, or the quasi parton distributions can be factorized with twist-2 parton distributions in the limit. 
This proposal has stimulated intensive studies of calculations of parton distributions,  a review about current progresses of studies in this field can be found in \cite{LAM}. 
 
Twist-2 parton distributions are defined with operators sandwiched between the same hadron state.
With different hadron states but the same operators one can define Generalized Parton Distributions(GPDs)  as introduced in \cite{DMGPD,DVCSJi}. GPDs contain more information than twist-2 parton distributions. It has been shown \cite{DVCSJi} one can obtain the quark- or gluon contributions to the proton spin from quark- or gluon 
twist-2 GPDs, respectively. The properties 
of twist-2 GPDs have been studied extensively and can be found in reviews \cite{DMGPDR,MDI,BeRa}.
    
\par 
The idea with quasi PDFs can be generalized to the case of GPDs\cite{Ji}, where one defines quasi GPDs 
which can be calculated with lattice QCD. A lattice study of quasi GPDs can be found in \cite{CLZ,LGPD,Lin1,Lin2}. The defined quasi GPDs are related to twist-2 GPDs through factorization relations in the limit of large hadron momenta, in which quasi GPDs are given as convolutions 
of GPDs with perturbative coefficient functions. In the limit, the quasi GPDs contain the same nonperturbative physics 
as GPDs do.  In this work we study these factorization relations of quasi quark GPDs.

\par 
The factorization relations of quasi quark GPDs of a proton or a spin-1/2 hadron have been studied in \cite{MGPD1, MGPD2, MGPD3}. 
In \cite{MGPD1,MGPD2} one calculates quasi GPDs and GPDs at one-loop with quark states instead of hadron states. 
With such one-loop calculations, one can not obtain factorization relations of all quasi quark GPDs at one-loop. 
This can be explained with the case of unpolarized quark GPDs. With a twist-2 
operator two unpolarized quark GPDs can be defined. Similarly, one can define 
two quasi quark GPDs with a operator corresponding to the twist-2 
operator. Calculating these GPDs of quark states, one finds that one of the two GPDs and the corresponding quasi one is zero at tree-level. They are nonzero at one-loop. Therefore, the tree-level factorization relation of this GPD 
is obtained through one-loop calculations. For obtaining it at one-loop, one needs a two-loop calculation of the GPD with quark states.  In \cite{MGPD3} it is shown with operator product expansion that in fact the factorization relation of the two unpolarized GPDs has the same perturbative coefficient function. The same also applies for the polarized quasi GPDs. However, the perturbative coefficient functions in factorization relations in \cite{MGPD3} 
are still extracted from results of GPDs and quasi GPDs of quark states at one-loop.  

\par 
In this work we employ a different approach to study the factorization. The approach is based on diagram 
expansion which has been successfully used for the analysis of power corrections in DIS in \cite{EFP, JWQ}.  In this approach one directly calculates perturbative coefficient functions in the factorization without calculating individual GPD and individual quasi GPD of quark states.  With one-loop calculations we 
can derive factorization relations of all quasi GPDs at one-loop. From our study,  the factorization relations are between matrix elements used to define quasi GPDs and those used to define twist-2 GPDs. Individual quasi GPDs defined with one operator have the same factorization relation to the corresponding twist-2 GPDs, the reason for this is that the factorization or matching is between operators. Our factorization results apply not only to the case of a spin-1/2 hadron 
but also to the case of hadrons with  other spins.  This is like QCD factorization with parton distributions 
for a real scattering process with an initial hadron, where the perturbative coefficient functions do not depend
on the type of the initial hadron. In this work, we focus on factorizations of quasi quark GPDs, where the contributions of twist-2 gluon GPDs in the factorization are also derived.  The study of the factorization of quasi gluon GPDs is under progress.   

\par 
In the limit where the different hadron states of GPDs are taken the same, GPDs become PDFs. From our 
results we obtain the factorization relations between quasi PDFs and twist-2 PDFs, which have been studied extensively in \cite{M1,M2,WZZ,STYZ,IJJSZ,WZZZ,CWZ,LMQ}.  
Besides quasi PDFs one may also use the concept of pseudo PDFs in \cite{AnRa1} for the purpose to calculate PDFs. It has been shown that both quasi PDFs and pseudo PDFs capture the same physics 
at long distance in \cite{IJJSZ}. One can also define pseudo GPDs similarly to pseudo PDFs. The factorization of pseudo GPDs has been studied in \cite{AnRa2}. It is expected that quasi GPDs 
and pseudo GPDs have the same nonperturbative physics in analogy to the case of PDFs.

\par

Our work is organized as in the following: In Sect.2 we give definitions of twist-2 quark GPDs and corresponding quasi GPDs.  The factorization relations at tree-level are derived. In Sect.3 we study the contributions from twist-2 quark GPDs to the relations. In Sect.4 the contributions from twist-2 gluon GPDs 
are derived. In Sect. 5 we present our complete results of the factorization relations. A comparison with 
existing results is made.  Then we take the kinematical limit to obtain the factorization relations between quark quasi PDFs and twist-2 parton PDFs. 
Sect. 6 is our summary. 

\par\vskip20pt
\par
\noindent
{\bf 2.  Definitions and Tree-Level Results}

To give the definitions of quark GPDs we introduce the  light-cone coordinate system, in which a vector
$a^\mu$ is expressed as $a^\mu = (a^+, a^-, \vec a_\perp) =((a^0+a^3)/\sqrt{2}, (a^0-a^3)/\sqrt{2}, a^1, a^2)$ and $a_\perp^2
=(a^1)^2+(a^2)^2$. Two light-cone vector
$l^\mu=(1,0,0,0)$ and $n^\mu=(0,1,0,0)$ are introduced in the system.
\par

We consider a single hadron in the final- and initial state. The hadron has the momentum $p$ in the initial state and $p'$ in the final state, respectively. The initial- or final hadron moves closely along light-cone direction $l$, i.e., the $z$-components of the three momenta are large. The transverse momenta are small.   We introduce the following quantities
\begin{equation}
  P^\mu = \frac{1}{2}( p^\mu + p'^{\mu}) , \quad  \Delta^\mu = p'^{\mu} -p^\mu, \quad t =\Delta^2, 
  \quad \xi =  - \frac{\Delta^+} {2 P^+} =\frac{ p^{+} -p'^+} {p^{+ } + p'^+ },   
\end{equation}  
 and the gauge link: 
\begin{equation} 
   {\mathcal L}_n (y,\infty)  =  P\exp\biggr \{ -i g_s \int_0^\infty d \xi n\cdot G(\xi n + y) \biggr \}. 
\end{equation}  
We take a coordinate system in which $P^\mu$ is given by $P^\mu =(P^+,P^-,0,0)$.    
The twist-2 quark GPDs are defined and parameterized for a spin-1/2 hadron as: 
\begin{eqnarray}
 F_q (x,\xi, t) &=& \frac{1}{2}  \int \frac{d\lambda}{2\pi} 
                e ^{i x  P^+ \lambda }
                \langle p'  \vert
                \bar\psi (-\frac{\lambda}{2} n )  {\mathcal L}_n^\dagger (-\frac{\lambda}{2},\infty) \gamma^+ {\mathcal L}_n  (\frac{\lambda}{2},\infty)
                  \psi(\frac{\lambda}{2}  n)\vert p \rangle 
\nonumber\\
         &=& \frac{1}{2 P^+} \bar u(p') \biggr [ \gamma^+ H_q (x,\xi, t) + i \frac{ \sigma^{+\nu} \Delta_\nu}{2 m} 
            E_q (x,\xi,t) \biggr ] u(p),  
\nonumber\\
   F_{qL}  (x,\xi, t) &=& \frac{1}{2}  \int \frac{d\lambda}{2\pi} 
                e ^{i x  P^+ \lambda }
                \langle p'  \vert
                \bar\psi (-\frac{\lambda}{2} n )  {\mathcal L}_n^\dagger (-\frac{\lambda}{2},\infty) \gamma^+ \gamma_5 {\mathcal L}_n  (\frac{\lambda}{2},\infty)
                  \psi(\frac{\lambda}{2}  n)\vert p \rangle
\nonumber\\
         &=& \frac{1}{2 P^+} \bar u(p') \biggr [ \gamma^+\gamma_5   H_{qL} (x,\xi, t) +  \frac{ \gamma_5 \Delta^+   }{2 m} 
             E_{qL} (x,\xi,t) \biggr ] u(p)   
\nonumber\\
 F_{qT}^\nu  (x,\xi, t) &=& \frac{1}{2}  \int \frac{d\lambda}{2\pi} 
                e ^{i x  P^+ \lambda }
                \langle p'  \vert
                \bar\psi (-\frac{\lambda}{2} n )  {\mathcal L}_n^\dagger (-\frac{\lambda}{2},\infty) i\sigma^{+ \nu} {\mathcal L}_n  (\frac{\lambda}{2},\infty)
                  \psi(\frac{\lambda}{2}  n)\vert p \rangle 
\nonumber\\
         &=& \frac{1}{2 P^+} \bar u(p') \biggr [ i\sigma^{+\nu}  H_{qT} (x,\xi, t)  
            + \frac{1}{m^2} ( P^+ \Delta^\nu - \Delta^+ P^\nu) \tilde H_{qT}(x,\xi,t) 
\nonumber\\             
       && + \frac{1}{2 m} ( \gamma^+ \Delta^\nu - \Delta^+ \gamma^\nu) E_{qT} (x,\xi,t) 
       + \frac{1}{ m} ( \gamma^+ P^\nu - P ^+ \gamma^\nu) \tilde E_{qT} (x,\xi,t)
           \biggr ] u(p), 
\label{DGPD}                                                    
\end{eqnarray}
where $\nu$ is transverse.  With the three matrix elements $F_q, F_{qL}$ and $F_{qT}$ one defines 
unpolarized-, longitudinally polarized and transversity quark GPDs respectively.  There are in total 8 quark GPDs of a spin-1/2 hadron. 
They have the support $-1\leq x \leq 1$ with $x$ as a parton momentum fraction.  The parameter $\xi$ is restricted in the region with $\vert \xi\vert \leq 1$.  
Later, we will introduce gluon GPDs.   

From the definitions, the operators used to define quark GPDs depend on time. Therefore, they can not be calculated directly with Lattice QCD because Lattice QCD is formulated in Euclidian space.  
However, one can introduce quasi GPDs and calculate them directly with Lattice QCD. The obtained  
quasi GPDs can be used to determine GPDs.  
To introduce quasi GPDs, we use cartesian coordinate system and introduce a vector
$n_z^\mu =(0,0,0,-1)$ pointing the $-z$-direction. We define quasi-GPDs and give the parametrization for a spin-1/2 hadron:
\begin{eqnarray}
 {\mathcal F} _q (z,\xi, t) &=& \frac{1}{2}  \int \frac{d\lambda  }{2\pi} 
                e ^{  i  z \lambda  P^z    }
                \langle p'  \vert
                \bar\psi (-  \frac{\lambda}{2} n_z )  {\mathcal L}_z^\dagger (- \frac{\lambda}{2} n_z) \gamma^z {\mathcal L}_z  (\frac{\lambda}{2}n_z)
                  \psi( \frac{\lambda}{2}  n_z)\vert p \rangle 
\nonumber\\
         &=& \frac{1}{2 P^z} \bar u(p') \biggr [ \gamma^z {\mathcal H}_q (z,\xi, t) + i \frac{ \sigma^{z\nu} \Delta_\nu}{2 m} 
           {\mathcal E}_q (z,\xi,t) \biggr ] u(p),  
\nonumber\\
  {\mathcal F} _{qL}  (z,\xi, t) &=& \frac{1}{2}  \int \frac{d\lambda }{2\pi} 
                e ^{  i  z \lambda P^z    }
                \langle p'  \vert
                \bar\psi ( -\frac{\lambda }{2} n_z )  {\mathcal L}_z^\dagger (- \frac{\lambda }{2} n_z) \gamma^z \gamma_5 {\mathcal L}_z  ( \frac{\lambda }{2}n_z)
                  \psi( \frac{\lambda }{2}  n_z)\vert p \rangle 
\nonumber\\
         &=& \frac{1}{2 P^z} \bar u(p') \biggr [ \gamma^z\gamma_5   {\mathcal H} _{qL}  (z,\xi, t) +  \frac{ \gamma_5 \Delta^z   }{2 m} 
             {\mathcal E} _{qL} (z,\xi,t) \biggr ] u(p)   
\nonumber\\
 {\mathcal F}^\nu _{qT}  (z,\xi, t) &=& \frac{1}{2}  \int \frac{d\lambda }{2\pi} 
                e ^{  i z\lambda  P^z    }
                \langle p'  \vert
                \bar\psi ( -\frac{\lambda }{2} n_z )  {\mathcal L}_z^\dagger (- \frac{\lambda }{2} n_z) i\sigma^{z \nu}  {\mathcal L}_z  ( \frac{\lambda }{2}n_z )
                  \psi( \frac{\lambda }{2}  n_z)\vert p \rangle 
\nonumber\\
         &=& \frac{1}{2 P^z} \bar u(p') \biggr [ i\sigma^{z\nu}  {\mathcal H} _{qT} (z,\xi, t)  
            + \frac{1}{m^2} ( P^z \Delta^\nu - \Delta^z P^\nu) \tilde {\mathcal H}_{qT}(z,\xi,t) 
\nonumber\\             
       && + \frac{1}{2 m} ( \gamma^z \Delta^\nu - \Delta^z \gamma^\nu) {\mathcal E} _{qT}(z,\xi,t) 
       + \frac{1}{ m} ( \gamma^z P^\nu - P ^z \gamma^\nu) \tilde {\mathcal E}_{qT} (z,\xi,t)
           \biggr ] u(p), 
\label{DQGPD} 
\end{eqnarray}
where the gauge link is along the $z$-direction:
\begin{eqnarray}
{\mathcal L}_z (y) = P\exp\biggr \{ -i g_s \int_0^\infty d \xi n_z\cdot G(\xi n_z + y) \biggr \}.                   
\end{eqnarray}
The defined quasi GPDs do not depend on time. $z$ is the momentum fraction which is unrestricted, i.e., 
$-\infty < z <\infty$.

The introduced  quasi quark GPDs can be related to GPDs. In the limit of large $P^z$, quasi GPDs can be factorized with GPDs, where perturbative coefficient functions are free from any soft divergences and 
do not depend on the small $t$. 
  At the leading power of $P^z$ only twist-2 GPDs are involved in the factorization.  
At tree-level, the factorization with twist-2 GPDs or matching to twist-2 GPDs can be shown with operator 
product expansion by the comparison  of moments of quasi GPDs and the corresponding GPDs  in the limit of large $P^z$.  In our work we will  use diagram expansion for doing this. 
This approach has been successfully used for the analysis of power corrections in DIS in \cite{EFP, JWQ}.

\begin{figure}[hbt]
\begin{center}
\includegraphics[width=16cm]{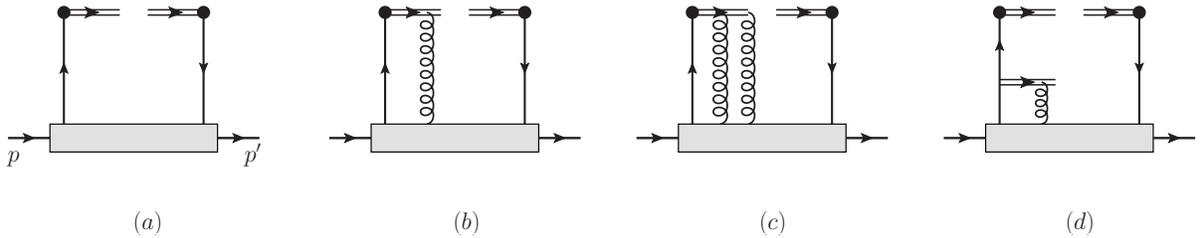}
\end{center}
\caption{The tree-level diagrams for quasi-GPDs. The double lines in the top of diagrams stand for gauge links along the $z$-direction. In (d) the double line in the middle stands for the gauge link along the $n$-direction(see the corresponding text after Eq.(10)).       } 
\label{Fig1}
\end{figure}

The tree-level factorization is given by diagrams in Fig.1. 
We take ${\mathcal F}_q$ as an example to explain the approach. 
We first consider the contribution from Fig.1a. The contribution can be written as:
\begin{eqnarray} 
 {\mathcal F}_q(z, \xi, t) ) \biggr\vert_{1a}  = \int d^4 k \delta ( z P^z -k^z) \int \frac{ d^4 y }{2 (2\pi)^4} e^{i y\cdot k}   
  \langle p'  \vert
                \bar\psi (-\frac{y}{2}  )    \gamma^z 
                  \psi( \frac{y}{2}  )\vert p \rangle, 
\end{eqnarray}                   
where the left quark-line carries the momentum $k-\frac{1}{2} \Delta$, the right quark-line carries the  momentum $k+ \frac{1}{2} \Delta$.  GPDs are used to describe a class of exclusive forward scattering 
processes, like deeply virtual Compton scattering in forward region. Therefore, as said before, 
the momentum $p$ and $p'$ have the largest component in the $+$-direction. This results in that 
the momentum $k$ and the difference $\Delta^\mu$ scales
\begin{equation} 
     k^\mu =(k^+,k^-,k^1,k^2) \sim (1,\lambda^2,\lambda,\lambda), \quad 
     \Delta^\mu =(\Delta^+,\Delta^-,\Delta^1,\Delta^2) \sim (1,\lambda^2,\lambda,\lambda)    
\end{equation} 
with $\lambda\ll 1$. Here $\lambda$ is either proportional to $\Lambda_{QCD}/Q$ or to $\vert \Delta_\perp\vert/Q$. $Q$ is the large scale $P^z =n_z\cdot P$. 
At the leading power we can neglect all contributions suppressed by powers of $\lambda$.        
\par
It is noted that the $z$-component of $P^\mu$ and $\gamma^\mu$ is related to the components in the light-cone coordinate system as:
\begin{equation} 
    P^z = \frac{1}{\sqrt{2}} ( P^+ - P^-), \quad \gamma^z =  \frac{1}{\sqrt{2}} ( \gamma^+ - \gamma^-).     
\end{equation}
The limit of large $P^z$ is equivalent to the limit of large $P^+$. In the limit $P^-$ is small and can be neglected.
The effect of $\gamma^-$ in the relation of $\gamma^z$ is a twist-4 effect, which is power-suppressed and again can be neglected. 
Keeping only the leading power contribution, we obtain from Fig.1a:     
\begin{eqnarray} 
 {\mathcal F}_ q(z, \xi, t) \biggr\vert_{1a}  = \int \frac{ d y^-  }{2 (2\pi)} e^{i y^- z P^+ }   
   \langle p'  \vert
                \bar\psi (-\frac{y^- n }{2}  )    \gamma^+ 
                  \psi( \frac{y^- n }{2}  )\vert p \rangle  +\cdots, 
\end{eqnarray}  
where $\cdots$ stand for power-suppressed contributions. The matrix element here consists  
only of quark fields. It is not exactly the same as that used to define the corresponding GPD. 
The contributions  from Fig.1b, 1c consist of one- and two gluon exchanges.  The contribution
from these diagrams is related to the matrix element containing one- and two gluon fields besides  quark fields, respectively.
In the considered case, there is a power counting for these gluon fields in Feynman gauge:
\begin{equation} 
    G^\mu =(G^+,G^-,G^1,G^2) \sim (1,\lambda^2,\lambda,\lambda). 
\end{equation} 
With this power counting, the leading contributions come
from diagrams with exchanges of  $G^+$-gluons. The momenta carried by these gluon fields 
obey the same power counting. The leading contributions are given by keeping only the $+$-components of 
momenta. The leading power contribution of Fig.1b is then obtained by replacing 
the gauge link along the $n_z$-direction with the gauge link along the $n$-direction.  The contribution can be expressed with Fig.1d, where the gauge link attached by a gluon line is along the $n$-direction. It is noted that in Fig.1d the quark line above the gauge link along the $n$-direction is not a quark propagator.   
The leading contributions with exchange of one-, two- and more gluons can be easily summed into gauge links along the $n$-direction. The gauge link just takes   
the form as that in the definition of $F_q$. Therefore we have at tree-level and leading power:
\begin{equation} 
     {\mathcal F}_q (z,\xi, t) =  F_q(z, \xi, t) + \cdots.
\label{TFQ}      
\end{equation} 
 where $\cdots$ stand for power-suppressed contributions. Similarly, one has for other two quasi GPDs:
 \begin{equation} 
     {\mathcal F}_{qL} (z,\xi, t) =  F_{qL}(z, \xi, t),  \quad 
     {\mathcal F}_{qT} (z,\xi, t) =  F_{qT}(z, \xi, t), 
\label{TFQS}       
\end{equation} 
 where power corrections are neglected. Besides power corrections there are corrections beyond the 
 order of $\alpha_s^0$, which will be studied in the next sections.

\begin{figure}[hbt]
\begin{center}
\includegraphics[width=16cm]{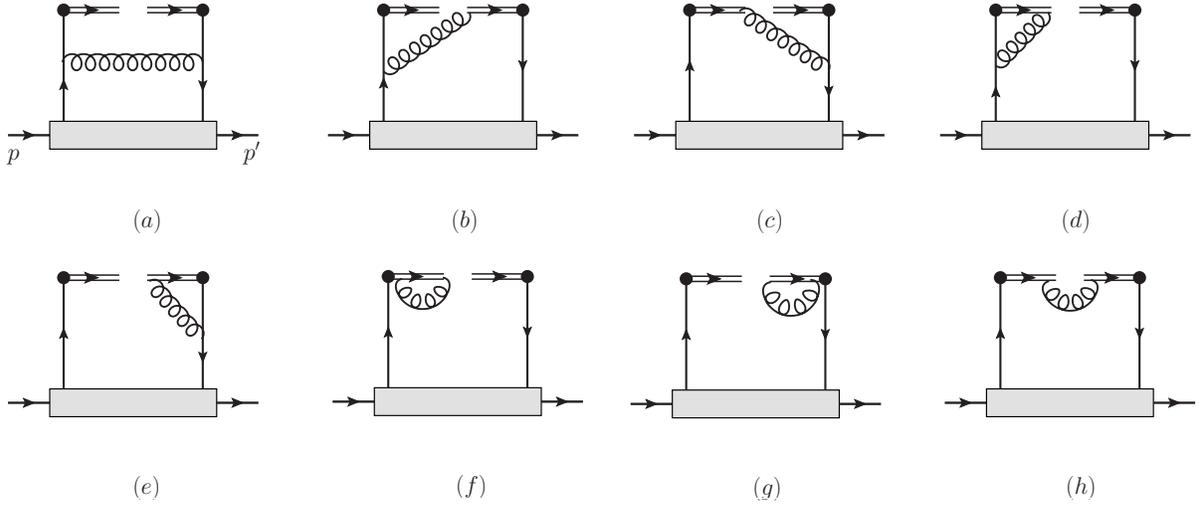}
\end{center}
\caption{The one-loop diagrams for quasi-GPDs. These diagrams represent contributions from quark GPDs.      } 
\label{Fig2}
\end{figure}

\par\vskip20pt

\noindent 
{\bf 3. One-Loop Factorization: Contributions from Quark GPDs} 

In this section we study the one-loop correction to the tree-level results in  the last section.
At one-loop the tree-level results receive corrections at the order of $\alpha_s$ from Fig.2, where double lines in the top of each diagram 
represent gauge links along the $n_z$-direction.  In this section we use Feynman gauge and dimensional regularization for U.V.- , I.R.- and collinear divergences.  
\par\vskip10pt
\noindent
{\bf 3.1 One Loop Contribution to quasi quark GPDs } 

We first take ${\mathcal F}_q$ to explain how to calculate the one-loop correction. The correction from Fig.2 can be written in the form: 
\begin{equation} 
{\mathcal F}_q^{(1)} (z,\xi, t) =  \int d^4 k   \Gamma_{ji, kl} (z, k, \Delta)  \left ( \gamma^z \right)_{kl}  \int \frac{ d^4 y }{2 (2\pi)^4} e^{i y\cdot k}   
  \langle p'  \vert
                \bar\psi_j (-\frac{y}{2}  )    
                  \psi_i( \frac{y}{2}  )\vert p \rangle,  
\end{equation}   
where the $\gamma$-matrix $\gamma^z$ stands for the projection of the contribution of ${\mathcal F}_q$  from the definition. $\Gamma$ is determined by the upper parts of diagrams, which can be written down with 
Feynman rule of QCD.  E.g, $\Gamma$ from Fig.2a reads: 
\begin{eqnarray} 
  \Gamma_{ji, kl}  (z, k,\Delta)\biggr\vert_{2a}   &=&   \int \frac{ d^4 k_g}{(2\pi)^4} \frac{-i g^{\mu\nu} }{k_g^2 +i\varepsilon} \delta ( k^z -k_g^z - z P^z) \biggr [ (-i g_s \gamma_\mu T^a) \frac{ i \gamma\cdot (k_2-k_g)}{(k_2-k_g)^2 + i\varepsilon}  \biggr]_{jk} 
\nonumber\\  
     && \biggr [ \frac{ i \gamma\cdot (k_1-k_g)}{(k_1-k_g)^2 + i\varepsilon} (-i g_s \gamma_\nu T^a) \biggr ]_{li} 
\end{eqnarray}  
with the momenta: 
\begin{equation} 
   k_1 = k -\frac{1}{2}  \Delta, \quad k_2 =  k + \frac{1}{2}  \Delta.  
 \end{equation}    
To obtain the leading- or twist-2 contributions, we can first neglect all components of $k$ and $\Delta$ in $\Gamma$ except the $+$-components,  according to the power counting stated in the last section. 
After doing this, we have
\begin{eqnarray} 
{\mathcal F}_q^{(1)}  (z,\xi, t) =  \int d k^+    \Gamma_{ji, kl} (z,  \hat k, \hat \Delta)  \left ( \gamma^z \right)_{kl}  \int \frac{ d  y^-  }{4\pi } e^{i y^-  k^+ }   
  \langle p'  \vert
                \bar\psi_j (-\frac{y^-}{2} n   )    
                  \psi_i( \frac{y^-}{2} n  )\vert p \rangle, 
\end{eqnarray} 
with 
\begin{equation}                    
 \hat k^\mu = (k^+, 0,0,0) = (xP^+, 0,0,0) , \quad \hat\Delta^\mu =(\Delta^+, 0,0,0) = (-2\xi P^+,0,0,0).                   
\end{equation}   
The hadronic matrix elements with different indices do not contribute at the same order of power. At leading power only those matrix elements give the leading power- or twist-2 contributions:  
\begin{eqnarray}   
\int \frac{ d  y^-  }{4\pi } e^{i y^-  k^+ }   
  \langle p'  \vert
                \bar\psi_j (-\frac{y^-}{2} n   )    
                  \psi_i( \frac{y^-}{2} n  )\vert p \rangle &=& \frac{1}{4 N_c} \biggr [ 
                     \gamma^-  F_q (x,\xi,t) + 
                   \gamma_5 \gamma^- F_{qL}  (x,\xi,t)  
 \nonumber\\                  
                  && + g_{\perp\mu\nu}   i\sigma^{\mu -}  F_{qT}^\nu  (x,\xi,t)
                   \biggr ]_{ij}  + \cdots,  
\end{eqnarray} 
where $\cdots$ denote higher-twist contributions. It is noted that in the expansion with $\gamma$-matrices of the 
matrix element in the left-hand-side the coefficient functions are quark GPDs defined in Eq.(\ref{DGPD}) but without 
gauge links. We have added the gauge links to write the coefficient functions as quark GPDs. This will not affect  our results at one-loop. 
For ${\mathcal F}_q$,  only the matrix element with the structure $\gamma^-$ gives the contribution. 
Therefore, we have: 
\begin{eqnarray} 
{\mathcal F}_q^{(1)}  (z,\xi, t) =   \int d k^+   {\mathcal M} (z, \hat k, \hat \Delta )  F_q (x, \xi,t) 
\label{FQ}                            
\end{eqnarray} 
with $k^+ = x P^+$ and   
\begin{equation} 
      {\mathcal M} (z, \hat k, \hat \Delta ) =\frac{1}{4 N_c}  \gamma^-_{ij} \Gamma_{ji, kl} (z, \hat k, \hat \Delta)  \left ( \gamma^z \right)_{kl}. 
\end{equation}
Similarly, one can work out the one-loop correction to other two quasi quark GPDs as:
\begin{eqnarray} 
{\mathcal F}_{qL}  (z,\xi, t)  &=&  \int d k^+    {\mathcal M}_L  (z, \hat k, \hat \Delta ) F_{qL} (x,\xi,t), 
\nonumber\\
 {\mathcal F}^\mu_{qT}  (z,\xi, t)  &=&   g_{\perp\nu\rho} \int d k^+   {\mathcal M}_T^{\mu\nu}  (z, \hat k, \hat \Delta )  F_{qT} ^\rho (x,\xi,t), 
\nonumber\\
 {\mathcal M}_L  (z, \hat k, \hat \Delta ) & = - & \frac{1}{4 N_c} \left (  \gamma^-\gamma_5\right ) _{ij} \Gamma_{ji, kl} (z, \hat k, \hat \Delta)  \left ( \gamma^z \gamma_5 \right)_{kl} , 
\nonumber\\
 {\mathcal M}_T^{\mu\nu}  (z, \hat k, \hat \Delta )  &=&  -  \frac{1}{4 N_c}  \left (  i \sigma^{-\nu } \right ) _{ij} \Gamma_{ji, kl} ( z, \hat k, \hat \Delta)  \left ( i\sigma^{z\mu}  \right)_{kl}.  
\end{eqnarray} 
There is no mixing between contributions from operators with different $\gamma$-matrices. 

It is rather straightforward to calculate contributions from each diagram to all ${\mathcal M}$'s. We will use dimensional regularization with the space-time dimension $d=4-\epsilon$ to regularize all divergences. 
It is noted that we have here only massless loop integrals. We list first the results of ${\mathcal M}$
for the correction to ${\mathcal F}_q$. 
The contributions from first 5 diagrams are: 
\begin{eqnarray} 
   {\mathcal M} \biggr\vert_{2a} &=& \frac{\alpha_s C_F }{2\pi \sqrt{2} P^z} \biggr \{  \biggr (-\frac{2}{\epsilon_c} -\ln\frac{4\pi \mu_c^2}{ (2P^z)^2  e^\gamma} +1\biggr ) \biggr [ 
   \frac{z+\xi}{4\xi(x+\xi)} \epsilon(z+\xi) -\frac{z-\xi}{4\xi(x-\xi)} \epsilon(z-\xi) 
\nonumber\\   
     &&   +\frac{x-z}{2(x^2-\xi^2) } \epsilon(x-z) 
    \biggr ] 
 + \frac{z+\xi}{4 \xi (x+\xi) }  \epsilon(z+\xi) \ln (z+\xi)^2  - \frac{z-\xi}{4\xi (x-\xi)}\epsilon(z-\xi) \ln (z-\xi)^2  
 \nonumber\\
     && +\frac{x-z}{2 (x^2-\xi^2)} \epsilon(x-z) \ln (x-z)^2\biggr \},    
 \nonumber\\
      {\mathcal M} \biggr\vert_{2b} &=& \frac{\alpha_s C_F }{4\pi \sqrt{2} P^z} \frac{z+\xi }{(x-z)(x+\xi)}
     \biggr [ \biggr (-\frac{2}{\epsilon_c} -\ln\frac{4\pi \mu_c^2}{ (2 P^z)^2  e^\gamma} \biggr ) \biggr ( \epsilon(z+\xi) +\epsilon(x-z) \biggr )  
\nonumber\\     
    &&  
 + \epsilon(z+\xi) (\ln (z+\xi)^2-1 ) + \epsilon(x-z) \biggr (  \ln (x-z)^2 + \frac{x- z}{z+\xi}  \biggr ) \biggr ], 
 \nonumber\\
    {\mathcal M} \biggr\vert_{2c} &=&  {\mathcal M} \biggr\vert_{2b} (\xi \to -\xi),   
\nonumber\\
     {\mathcal M} \biggr\vert_{2d} &=& \delta (z-x) \frac{\alpha_s C_F }{4\pi  \sqrt{2} P^z } \biggr\{ 
       -2 -2 \ln^2 \vert x+\xi \vert +\ln^2 (1+x) +\ln^2 (1-x) +\frac{1}{3} \pi^2  + \ln\frac{(x+\xi)^2\mu^2} {(2 P^z)^2}
\nonumber\\
  &&   + \biggr (\frac{2}{\epsilon_c}  + \ln\frac{4\pi \mu_c^2}{(2 P^z)^2  e^\gamma} \biggr)    \biggr [ \ln \frac{(x+\xi)^2}{1-x^2} -2 + \int_{-1} ^1    dy  \frac{\epsilon(x-y) }{(x-y )}    \biggr] 
   - \int_{-1} ^1    dy  \frac{\epsilon(x-y) }{(x-y )} \ln(x-y)^2  \biggr \},               
 \nonumber\\
    {\mathcal M} \biggr\vert_{2e} &=&  {\mathcal M} \biggr\vert_{2d} (\xi \to -\xi).           
\label{GFIG2A}       
\end{eqnarray}   
In the above, the collinear divergences are represented by the poles $1/\epsilon_c$ and $\mu_c$ is the corresponding 
scale. There is an U.V. divergence in the contribution from Fig.2d,  the corresponding U.V. pole is subtracted. $\mu$ is the U.V. renormalization scale. The function $\epsilon (x)$ is the sign function defined 
as
\begin{equation} 
  \epsilon(x) = \theta(x)-\theta(-x). 
\end{equation}    
Besides divergences regularized with dimensional regularization, there is an end-point divergence in the contribution from Fig.2b at $x=z$. In the integral in the contribution from Fig.2d, there is an end-point divergence too. Taking these contributions as distributions, the sum of contributions from Fig.2b and Fig.2d is free from such divergences: 
\begin{eqnarray} 
     {\mathcal M} \biggr\vert_{2b+2d} &=& 
     \frac{\alpha_s C_F }{4\pi \sqrt{2} P^z  } \biggr \{ 
       \biggr ( - \frac{2}{\epsilon_c} - \ln\frac{4\pi \mu_c^2}{ (2 P^z)^2  e^\gamma} \biggr ) \biggr [  \biggr ( \frac{\epsilon(x-z)}{(x-z)} \biggr)_+  -\frac{\epsilon(x-z)}{x+\xi} + \frac{ (z+\xi) \epsilon(z+\xi )  }{(x-z)(x+\xi)} 
\nonumber\\       
       && + \delta(x-z) \biggr ( 2-  \ln\frac{(x+\xi)^2}{1-x^2} \biggr )  \biggr ]  +  \biggr ( \frac{ \epsilon (x-z) \ln (x-z)^2 }{x-z} \biggr )_+ 
 \nonumber\\
  && +\frac{(z+\xi)  \epsilon(z+\xi) }{( x-z)(x+\xi) } \biggr (   \ln (z+\xi)^2-1  \biggr )  + \frac{\epsilon(x-z)}{x+\xi} 
     -\frac{ \ln (x-z)^2 }{x+\xi}\epsilon(x-z)         
\nonumber\\
   &&  +\delta (x-z) \biggr [   \ln\frac{(x+\xi)^2\mu^2} {(2 P^z)^2}  -2  -2 \ln^2 \vert x+\xi \vert  + \ln^2 (1+x) +\ln^2 (1-x) +\frac{1}{3} \pi^2 \biggr ] \biggr \},           
\end{eqnarray}   
where the $+$-distribution is defined as:
\begin{equation} 
   \int_{-1}^1 dx   \biggr [ \frac{\epsilon (x-z) f(x)}{x-z}\biggr ]_+ t (x)  =  \int_{-1}^1 dx   \biggr \{ \biggr [ \frac{\epsilon(x-z) f(x)}{x-z}\biggr ]  -  \delta (x-z) \int_{-1}^1 dy \biggr [ \frac{\epsilon(y-z) f(y)}{y-z}\biggr ]     \biggr \}  t(x), 
\end{equation} 
with $t(x)$ as a test function.  Among the contributions from the last three diagrams in Fig.2,  there are U.V. logarithm divergences represented by poles of $1/\epsilon$ from Fig.2f and Fig.2g, which are subtracted. There are also power divergences represented by poles of $1/(1-\epsilon)$. These divergences are taken to be finite in $\overline{\rm MS}$-scheme used in this work.  The subtraction of similar power divergences in quasi PDFs has been discussed in \cite{IMQY}.  There are end-point divergences from the three diagrams. However, these divergences are cancelled in the sum.  We have the sum: 
\begin{eqnarray} 
{\mathcal M} \biggr\vert_{2h+2f +2g } &=& - \frac{\alpha_s C_F }{2\pi \sqrt{2} P^z } \biggr [ 
  \biggr (  \frac{\epsilon(x-z)}{( x-z) } \biggr )_+ + \delta (x-z) \biggr ( \ln (1-x^2) -  \ln \frac{\mu^2}{(2 P^z)^2} - 2 \biggr ) \biggr ].  
\end{eqnarray}      
There is no collinear divergence.  
 
The one-loop correction to ${\mathcal F}_q$ is the sum of contributions from Fig.2.  For one-loop corrections 
to other two quasi-GPDs calculations can be done in a similar way. However, such calculations are not needed, because one can find the following relations between one-loop corrections of the three quasi GPDs.  For the one-loop correction of ${\mathcal F}_{qL}$ it is easy to find: 
\begin{equation} 
   {\mathcal M}_L  (\hat k, \hat \Delta ) =  {\mathcal M}  (\hat k, \hat \Delta ), 
\label{ML}    
\end{equation} 
i.e., the contribution to ${\mathcal F}_{qL}$ is the same as that to ${\mathcal F}_{q}$.         
For ${\mathcal F}_{qT}$ we find that the there is no contribution from Fig.2a.  For contributions from other diagrams in Fig.2 
a simple relation can be found:
\begin{equation}  
{\mathcal M}_T^{\mu\nu}  (\hat k, \hat \Delta ) \biggr\vert_{2 i  } = g_\perp^{\mu\nu} {\mathcal M}  (\hat k, \hat \Delta ) \biggr\vert_{2 i  },  \quad {\rm for}\  i\neq a. 
\label{MT}
\end{equation} 
Therefore, we have in this section all results of one-loop corrections to the all three quasi GPDs in terms of corresponding GPDs. However, the results are not correct.  Before we discuss this in the next subsection, we discuss briefly the one-loop renormalization of quasi GPDs based on our results. 

Since we have U.V. divergences here from Fig.2, the renormalization of ${\mathcal  F}_q$ is not only determined by wave function renormalization in Feynman gauge but also by divergences from Fig.2. These divergences can be removed by introducing another renormalization constant $Z_{\mathcal F}$:
\begin{equation} 
   \biggr ( {\mathcal F}_q (z,\xi,t)  \biggr )_0 = Z_2 Z_{\mathcal F}{\mathcal F}_q (z,\xi,t,\mu ) +\cdots 
\label{UVQF}
\end{equation} 
where the quantity in the left side is unrenormalized one, and $\cdots$ is possible operator mixing term, 
which is zero at the order we consider according to our study in Sect. 3. 
 $Z_2$ is the wave function renormalization constant of quark field in Feynman gauge, which is 
\begin{equation} 
  Z_2 = 1- \frac{\alpha_s C_F }{4\pi} \biggr ( \frac{2}{\epsilon} -\gamma + \ln (4\pi) \biggr ) +\cdots, 
\end{equation} 
and $Z_{\mathcal F}$ is determined by the U.V. divergence from Fig.2:
\begin{equation} 
 Z_{\mathcal F} = 1 + \frac{\alpha_s C_F}{\pi}  \biggr ( \frac{2}{\epsilon} -\gamma + \ln (4\pi)  \biggr ) 
       +\cdots. 
\end{equation}         
With these results one can derive the evolution equation of the renormalization scale $\mu$:
\begin{equation} 
     \frac{d } {d \ln \mu^2 }{\mathcal F}_q(z,\xi,t,\mu)  = \frac{ 3 \alpha_s C_F}{4 \pi }{\mathcal F}_q(z,\xi,t,\mu)   + {\mathcal O}(\alpha_s^2). 
\label{AZC}      
\end{equation}      
This is expected from the wave function renormalization in the axial gauge in $\overline{\rm MS}$-scheme. 
The renormalization constant and the $\mu$-dependence given here agree with the existing results in \cite{AZC}.  
For other two quasi GPDs the renormalization is the same.

\par\vskip20pt
\noindent
{\bf 3.2. Subtraction  } 

The results of one-loop corrections given in terms of GPDs in the last subsection are not correct due to a double counting. 
This can be seen from the tree-level results in Sect.2 or tree-level diagrams in Fig.1.  
In Fig.1, if we pull quark lines from the grey boxes which represent quark GPDs, we will meet the same diagrams as given in Fig.2, 
but the gauge links in this case is along the light-cone $n$-direction instead of along the $n_z$-direction. 
With the light-cone $n$-direction in the diagrams, the exchanged gluons are essentially collinear gluons.
This implies that the one-loop correction of exchanges of collinear gluons is already included in GPDs in the tree-level 
results given in Eq.(\ref{TFQ}, \ref{TFQS}). Therefore, the contributions from exchanges of collinear gluons 
are double-counted in the results of one-loop corrections in the last subsection.  The correct results are obtained if we subtract the collinear contribution from the one-loop corrections.  If the factorization of quasi-GPDs with twist-2 GPDs holds, the one-loop corrections after the subtraction are finite.  

The collinear contributions which need to be subtracted can be calculated from diagrams given in Fig.2 
with the gauge links,  which 
are now along the light-cone direction $n$. The contributions can be written in the form, e.g., in the case 
of ${\mathcal F}_{q}$, as:
\begin{eqnarray} 
 F_q^{(1)}  (z,\xi, t) =  \int d k^+  M (\hat k, \hat \Delta )   \int \frac{ d  y^-  }{4\pi } e^{i y^-  k^+ }   
  \langle p'  \vert
                \bar\psi (-\frac{y^-}{2} n   )  \gamma^+   
                  \psi ( \frac{y^-}{2} n  )\vert p \rangle,            
\end{eqnarray} 
with 
\begin{equation} 
      M (\hat k, \hat \Delta ) =\frac{1}{4 N_c}  \gamma^-_{ij} \Gamma_{ji, kl} ( \hat k, \hat \Delta)  \left ( \gamma^+ \right)_{kl}. 
\end{equation} 
Here, $\Gamma$ is the same as introduced in the last subsection.  The calculation is straightforward. Since the gauge links considered here are along the $n$-direction, the last three diagrams give no contributions. 
The contributions from the first 5 diagrams are: 
\begin{eqnarray} 
   M\biggr\vert_{2a} &=& \frac{\alpha_s C_F }{2\pi P^+}   \biggr (-\frac{2}{\epsilon_c} -\ln\frac{4\pi \mu_c^2}{\mu^2 e^\gamma} \biggr )  \biggr [ 
   \frac{z+\xi}{4\xi(x+\xi)} \epsilon(z+\xi) -\frac{z-\xi}{4\xi(x-\xi)} \epsilon(z-\xi) 
   +\frac{x-z}{2(x^2-\xi^2) } \epsilon(x-z) 
    \biggr ],  
\nonumber\\
   M\biggr\vert_{2b} &=& \frac{\alpha_s C_F }{4\pi P^+}   \biggr (-\frac{2}{\epsilon_c} -\ln\frac{4\pi \mu_c^2}{\mu^2 e^\gamma} \biggr )\frac{z+\xi}{(x-z)(x+\xi) } \biggr [ \epsilon(z+\xi) +\epsilon(x-z) \biggr ] ,
 \nonumber\\
M\biggr\vert_{2c} &=&M\biggr\vert_{2b} (\xi\to -\xi), 
\nonumber\\
   M\biggr\vert_{2d}  &=&  - \delta (x-z)  \frac{\alpha_s C_F }{4\pi P^+}    \biggr (-\frac{2}{\epsilon_c} -\ln\frac{4\pi \mu_c^2}{\mu^2 e^\gamma} \biggr ) \biggr [  \int_{-1}^{1}  d y \frac{\epsilon(x-y)}{x-y }  
     +\ln \frac{(x+\xi)^2}{(1-x^2) }  -2 \biggr],  
\nonumber\\
M\biggr\vert_{2e} &=&M\biggr\vert_{2d} (\xi\to -\xi).
\label{SGPD}   
\end{eqnarray} 
There are U.V. divergences which are already subtracted in the above. 
Comparing with the results in Eq.(\ref{GFIG2A}), the collinear contribution there represented by terms with the pole $1/\epsilon_c$  in each diagram is exactly the same as that in the above, respectively.   At first look, it seems that 
the above collinear contributions are not zero for $\vert z\vert >1$. However, with the fact that  $\vert x\vert \le 1$ 
and $\vert \xi \vert \le 1$ one can show that the above contributions are zero in the case with $\vert z\vert >1$.

\par 
The correct one-loop corrections are obtained by subtracting the results in Sect.3.1 with the contribution in Eq.(\ref{SGPD}). 
The subtraction can be simply done by replacing the divergent factor in Eq.(\ref{GFIG2A}) by: 
\begin{equation} 
   \frac{2}{\epsilon_c} + \ln\frac{4\pi \mu_c^2 }{e^\gamma (2 P^z)^2} \to \ln\frac{\mu^2 }{ (2 P^z)^2}.    
\end{equation}
After the subtraction the one-loop correction to ${\mathcal F}_q$ is finite.    
For the one-loop corrections to $F_{qL, qT}$  there are same relations as for ${\mathcal F}_{qL, qT}$
as given in Eq.(\ref{ML},\ref{MT}), respectively. Therefore, one needs to do the same replacement 
for the subtraction. After the subtraction the corrections are finite. We will give the correct and complete 
results in Sect.5.

\par\vskip20pt

\begin{figure}[hbt]
\begin{center}
\includegraphics[width=16cm]{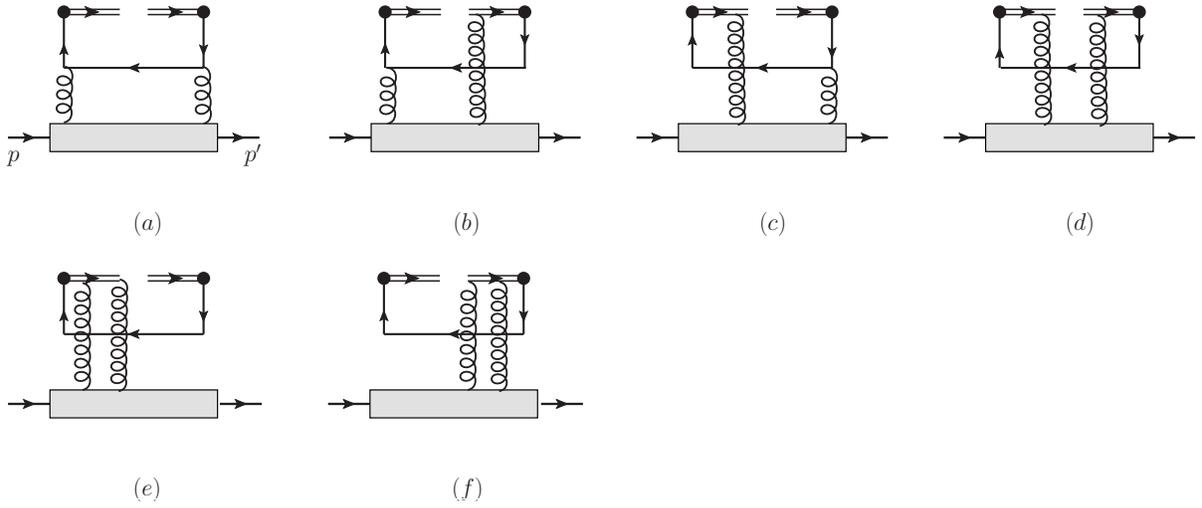}
\end{center}
\caption{The one-loop diagrams for quark quasi-GPDs. These diagrams represent contributions from gluon GPDs.      } 
\label{Fig3}
\end{figure}

\noindent 
{\bf 4. One-Loop Factorization: Contributions from Gluon GPDs}

Beyond tree-level, quasi quark GPDs receive contributions from gluon GPDs. The twist-2 gluon GPDs can be defined and parameterized for a spin-1/2 hadron  as: 
\begin{eqnarray} 
  F_g (x,\xi,t) &=& \frac{1}{ P^+} \int\frac{d\lambda}{2\pi} e^{ i x  P^+ \lambda} 
    \langle p' \vert g_{\perp\mu\nu} G^{+\mu} (-\frac{1}{2} \lambda n) G^{+\nu}  (\frac{1}{2} \lambda n ) \vert p \rangle  
\nonumber\\
     &=& -\frac{1}{2  P^+} \bar u(p') \biggr [ \gamma^+ H_g (x, \xi, t) + \frac{i\sigma^{+\alpha}} {2 m} \Delta_\alpha 
     E_g (x,\xi,t) \biggr ] u(p), 
\nonumber\\
  F_{gL}(x,\xi,t)     &=& \frac{i}{ P^+} \int\frac{d\lambda}{2\pi} e^{ i x  P^+ \lambda} 
    \langle p' \vert \epsilon_{\perp\mu\nu} G^{+\mu} (-\frac{1}{2} \lambda n) G^{+\nu} (\frac{1}{2} \lambda n ) \vert p \rangle
\nonumber\\
     &=& - \frac{1}{2  P^+} \bar u(p') \biggr [ \gamma^+ \gamma_5  H_{gL}  (x, \xi, t) + \frac{\gamma_5 \Delta^+ } {2 m} 
       E_{gL} (x,\xi,t) \biggr ] u(p),
\nonumber\\
  F^{\mu\nu }_{gT}(x,\xi,t)  &=&  \frac{1}{2  P^+} \int\frac{d\lambda}{2\pi} e^{ i x  P^+ \lambda} 
    \langle p' \vert  {\bf S} \biggr (   G^{+\mu } (-\frac{1}{2} \lambda n) G^{+\nu}  (\frac{1}{2}  \lambda n ) \biggr )   \vert p \rangle  
\nonumber\\
    &=& -\frac{1}{2 P^+} {\bf S} \biggr \{ \frac{ P^+ \Delta^\nu - \Delta^+ P^\nu}{2 m \bar P^+} \bar u(p') \biggr [ 
           i\sigma^{+\mu} H_{gT} (x,\xi,t) +\frac{ P^+ \Delta^\mu -\Delta^+ P^\mu}{m^2} \tilde H_{gT} (x,\xi,t) 
\nonumber\\
    && +\frac{\gamma^+ \Delta^\mu -\Delta^+ \gamma^\mu}{2 m} E_{gT} (x,\xi,t) + \frac{\gamma^+ P^\mu
       -P^+ \gamma^\mu }{m} 
     \tilde E_{gT} (x,\xi,t) \biggr ] u(p)  \biggr \},  
\label{GT2}                             
\end{eqnarray}  
where $\mu$ and $\nu$ are transverse. The notation ${\bf S}(\cdots)$ implies that the tensors in $(\cdots)$ are symmetric and traceless.  For simplicity we have omitted gauge links in $SU(N_c)$ adjoint representation
between operators of gluon field strength tensor.  There are in total eight twist-2 gluon GPDs for a spin-1/2 hadron. Their properties can be found in \cite{MDI, BeRa}.

At one-loop level, contributions are from 6 diagrams given in Fig.3.  We calculate these contributions in the light-cone gauge $n\cdot G = G^+=0$.  At leading power, the last five diagrams give no contribution. Nonzero contribution only comes from Fig.3a where the exchanged gluons are transversely polarized
and only carry momenta along the $+$-direction with the power counting discussed in Sect.2. 
The calculation is similar to those presented in the last section. We have the following results for ${\mathcal F}_q$: 
\begin{eqnarray} 
{\mathcal F}_q (z,\xi, t) \biggr\vert_{3a}       &=& - \frac{\alpha_s}{4\pi}  \int d k^+ 
\biggr \{  \biggr (-\frac{2}{\epsilon_c} -\ln\frac{4\pi \mu_c^2}{(2 P^z)^2 e^\gamma} \biggr ) \biggr [ 
    \frac{1}{4} \biggr ( 1 -2z \frac{ x -  z}{x^2-\xi^2 }\biggr ) \epsilon (x-z)
\nonumber\\
   && + \frac{\vert z+\xi \vert }{4\xi (x+\xi) } ( x+\xi-2z)      
 \biggr ] + \frac{\vert z+\xi \vert }{\xi(x+\xi)} \biggr ( z+ \frac{1}{4}(x+\xi-2z )\ln (z+\xi)^2 \biggr )   
\nonumber\\ 
  &&  + z\frac{x-z}{x^2-\xi^2} \epsilon(x-z)          
      + \frac{1}{4}  \epsilon(x-z)\biggr ( 1 -2z \frac{ x -  z}{x^2-\xi^2 }\biggr ) \ln (x-z)^2  + (\xi\to -\xi)    \biggr \}           
\nonumber\\
    && g_{\perp\mu\nu} \int \frac{ d y^-  }{2\pi} e^{i y^- k^+ }   
  \langle p'  \vert
                 G^{a,\mu}  (-\frac{y^- n }{2}  )    
                  G^{a,\nu}  (\frac{y^- n }{2}  )\vert p \rangle, 
\label{GluonG}                   
\end{eqnarray}  
and for ${\mathcal F}_{qL}$:
\begin{eqnarray}                   
{\mathcal F}_{qL}  (z,\xi, t) \biggr\vert_{3a}       &=& - \frac{\alpha_s}{4\pi}  \int d k^+ 
\biggr \{  \frac{1}{2} \biggr (\frac{2}{\epsilon_c} +\ln\frac{4\pi \mu_c^2}{(2 P^z)^2 e^\gamma} \biggr ) \biggr [ 
 ( x^2+\xi^2-2 x z) \biggr (\frac{\epsilon(z-\xi)}{2\xi (x-\xi)} 
 \nonumber\\
     && + \frac{\epsilon(x-z)}{2 ( x^2-\xi^2)} \biggr ) - \frac{x-z}{2 \xi}\epsilon(z-\xi)  \biggr ]  - \epsilon(x-z) \biggr ( \frac{x^2+\xi^2-2xz}{4(x^2-\xi^2)} \ln (x-z)^2 
\nonumber\\        
   && -  x \frac{x-z}{x^2-\xi^2} \biggr )+ \frac{1}{4\xi} \vert z+\xi\vert   \frac{\xi-x}{x+\xi} \biggr ( \ln (z+\xi)^2 - 2  \biggr ) 
      + (\xi\to -\xi)     
  \biggr \}           
\nonumber\\
    && i\epsilon_{\perp\mu\nu} \int \frac{ d y^-  }{2\pi} e^{i y^-  k^+ }   
  \langle p'  \vert
                 G^{a,\mu}  (-\frac{y^- n }{2}  )    
                  G^{a,\nu}  (\frac{y^- n }{2}  )\vert p \rangle.                                                           
\label{GluonGL}                                     
\end{eqnarray}   
It is noted that there is no U.V. divergence in the two contributions in the above. 
 
Because of the helicity conservation of perturbative QCD, ${\mathcal F}_{qT}$ receives no contribution from gluon GPDs. 
It is noted that in the light-cone gauge the matrix elements of gluon fields in the above are related 
to the matrix elements of gluon field strength tensors or twist-2 gluon GPDs as:  
\begin{eqnarray} 
 && \int \frac{ d y^-  }{2\pi} e^{i y^-  k^+ }   
  \langle p'  \vert
                 G^{a,\mu}  (-\frac{y^- n }{2}  )    
                  G^{a,\nu} ( \frac{y^- n }{2}  )\vert p \rangle 
\nonumber\\                  
                  && =  \frac{1}{ (x^2-\xi^2) (P^+)^2  } 
 \int \frac{ d y^-  }{2\pi} e^{i y^-  k^+ }   
  \langle p'  \vert
                 G^{a,+\mu}  (-\frac{y^- n }{2}  )    
                  G^{a,+\nu} ( \frac{y^- n }{2}  )\vert p \rangle,             
\end{eqnarray}
where $k^+$ is related to $P^+$ as $k^+ = xP^+$.  
\par 
For the same reasons discussed in Sect.3.2,  there is a double-counting in the contributions in Eq.(\ref{GluonG}, \ref{GluonGL}). A subtraction is needed for obtaining correct results. The contributions needed to be subtracted are given by the same diagrams in Fig.3 with the double lines representing gauge links along $n$-direction. In the light-cone gauge, the contributions only come from Fig.3a. They are: 
\begin{eqnarray}
 F_q^{(1)}  (z,\xi, t) \biggr\vert_{3a} &=& \frac{\alpha_s}{4\pi}  \biggr (-\frac{2}{\epsilon_c} -\ln\frac{4\pi \mu_c^2}{\mu^2 e^\gamma} \biggr )\int d k^+ \biggr [ 
    \frac{1}{2} \biggr ( 1 - 2z \frac{x-z}{x^2-\xi^2 }\biggr ) \epsilon (x-z)
\nonumber\\
   && + \frac{z+\xi}{4\xi (x+\xi) } ( x+\xi-2z) \epsilon(z+\xi) -\frac{z-\xi}{4\xi (x-\xi) } ( x-\xi-2z) \epsilon(z-\xi)     
 \biggr ]
\nonumber\\
   &&g_{\perp\mu\nu} \int \frac{ d y^-  }{2\pi} e^{i y^-  k^+ }   
  \langle p'  \vert
                 G^{a,\mu}  (-\frac{y^- n }{2}  )    
                  G^{a,\nu}  (\frac{y^- n }{2}  )\vert p \rangle      
\nonumber\\
 F_{qL}^{(1)}  (z,\xi, t) \biggr\vert_{3a} &=& \frac{ \alpha_s}{8\pi}    \biggr (-\frac{2}{\epsilon_c} -\ln\frac{4\pi \mu_c^2}{\mu^2 e^\gamma} \biggr )\int d k^+  \biggr [ 
 ( x^2+\xi^2-2 x z) \biggr (\frac{\epsilon(z-\xi)}{2\xi (x-\xi)} -\frac{\epsilon(z+\xi)}{2\xi (x+\xi) }
 \nonumber\\
     && + \frac{\epsilon(x-z)}{x^2-\xi^2} \biggr ) + \frac{x-z}{2 \xi} (\epsilon(z+\xi)-\epsilon(z-\xi) ) \biggr ] 
\nonumber\\
   && i \epsilon_{\perp\mu\nu} \int \frac{ d y^-  }{2\pi} e^{i y^-  k^+ }   
  \langle p'  \vert
                 G^{a,\mu}  (-\frac{y^- n }{2}  )    
                  G^{a,\nu} ( \frac{y^- n }{2}  )\vert p \rangle. 
\label{GSUB}                        
\end{eqnarray}        
Comparing the results in Eq.(\ref{GluonG},\ref{GluonGL}) with those in Eq.(\ref{GSUB}), they contain the same divergent contributions which are located in the region with $\vert z\vert \leq 1$. Therefore, after the subtraction, the correct one-loop corrections from twist-2 gluon GPDs are finite. The subtraction can be done with the same  replacement as discussed in Sect.3.2.

 \par\vskip20pt
 \noindent
 {\bf 5. Complete Results and the Limit of $\Delta^\mu\to 0$} 
 
 \par
 Based on results presented in previous sections, we can summarize our complete results of one-loop factorizations of quasi quark GPDs into the following form:  
 \begin{eqnarray} 
 {\mathcal F}_q (z,\xi,t) &=& F_q (z,\xi,t) +\frac{\alpha_s}{2\pi}  \int_{-1}^{1} dx \biggr [ 
    C_F h_q (x,\xi,z) F_q (x,\xi , t) + h_g (x,\xi,z) F_g (x,\xi,t) \biggr ] ,         
 \nonumber\\
{\mathcal F}_{qL}  (z,\xi,t) &=& F_{q L}  (z,\xi,t) +\frac{\alpha_s}{2\pi}  \int_{-1}^{1} dx \biggr [ 
    C_F h_{qL}  (x,\xi,z) F_q (x,\xi , t) + h_{gL}  (x,\xi,z) F_{gL}  (x,\xi,t) \biggr ] ,   
 \nonumber\\
{\mathcal F}_{qT}^\mu   (z,\xi,t) &=& F_{q T}^\mu  (z,\xi,t) +\frac{\alpha_s C_F}{2\pi}  \int_{-1}^{1} dx 
     h_{qT}  (x,\xi,z) F_{qT}^\mu  (x,\xi , t), 
\label{QFAC}      
 \end{eqnarray} 
 where $h$'s are perturbative coefficient functions. The functions related to twist-2 quark GPDs are:
 \begin{eqnarray} 
h_q (x,\xi,z) &=& \biggr ( \ln\frac{ (2P^z)^2}{\mu^2}  +1\biggr ) \biggr [ 
   \frac{\vert z+\xi \vert }{4\xi(x+\xi)}  -\frac{\vert z-\xi \vert }{4\xi(x-\xi)} +\frac{\vert x-z\vert }{2(x^2-\xi^2) } 
    \biggr ] + \frac{\vert z+\xi \vert }{4 \xi (x+\xi) }   \ln (z+\xi)^2  
\nonumber\\   
     &&      
   - \frac{\vert z-\xi \vert }{4\xi (x-\xi)} \ln (z-\xi)^2 +\frac{\vert x-z\vert }{2 (x^2-\xi^2)}  \ln (x-z)^2   
 + h_{qT}  (x,\xi,z), 
\nonumber\\
h_{qL}  (x,\xi,z) &=&h_q (x,\xi,z),          
\nonumber\\
h_{qT}  (x,\xi,z) &=&\frac{1}{2}   \biggr \{ 
         \biggr ( \frac{\epsilon(x-z)}{(x-z)}\ln\frac{ (2(x-z) P^z)^2} { e \mu^2}   \biggr)_+  -\frac{\epsilon(x-z)}{x+\xi}  \ln \frac{ (2 (x-z) P^z)^2}{e \mu^2}  
\nonumber\\         
  &&     
 +\frac{  \vert z+\xi  \vert }{( x-z)(x+\xi) }  \ln \frac{ (2(z+\xi) P^z)^2}{ e \mu^2}         
  +\delta (x-z) \biggr [  - \ln\frac{ (2 P^z)^2} {e \mu^2}\ln\frac{(x+\xi)^2}{1-x^2}\nonumber\\   
   &&      -2 \ln^2 \vert x+\xi \vert  + \ln^2 (1+x) 
 + \ln^2 (1-x)   +\frac{1}{3} \pi^2 \biggr ]  \biggr \}  +  (\xi \to -\xi)        
      .
\label{QCEQ}       
\end{eqnarray} 
The perturbative coefficient functions related to gluon GPDs are:
\begin{eqnarray} 
h_g (x,\xi, z)       &=&  \frac{1}{2 (x^2-\xi^2)}  
\biggr \{    \epsilon (x-z) \biggr (  
    \frac{x^2-\xi^2 +2 z^2-2xz}{4 (x^2-\xi^2) } \ln\frac{\mu^2}{(2 (x-z)  P^z)^2} - z\frac{x-z}{x^2-\xi^2} \biggr ) 
\nonumber\\    
 && + \epsilon(z+\xi)  \biggr ( \frac{z+\xi}{4\xi (x+\xi) } ( x+\xi-2z)  \ln\frac{\mu^2}{(2 (z+\xi) P^z)^2}  - z\frac{z+\xi}{\xi(x+\xi)} \biggr )  \biggr \}             
    +(\xi \to -\xi)  , 
\nonumber\\                   
h_{gL}  (x,\xi, z)      &=& - \frac{1}{2(x^2-\xi^2) }  
\biggr \{  \epsilon(x-z) \biggr (   
 \frac{ x^2+\xi^2-2 x z}{4 (x^2-\xi^2)}  \ln\frac{\mu^2}{(2 (x-z)  P^z)^2}  +x\frac{x-z}{x^2-\xi^2} \biggr ) 
 \nonumber\\
 &&  + \frac{(x-\xi) \vert z+\xi\vert }{4 \xi (x+\xi) }   \biggr (  \ln\frac{\mu^2}{(2 (z+\xi) P^z)^2}  + 2 \biggr )  \biggr \}  
      + (\xi\to -\xi)     .  
\label{QCEG}                                                                                                     
\end{eqnarray} 
From our results, the factorization relation of each individual quasi GPD in Eq.(\ref{DGPD}) with the corresponding  twist-2 quark GPD and corresponding twist-2 gluon GPD can be read from Eq.(\ref{QFAC},\ref{QCEQ},\ref{QCEG}). E.g.,  in the case of a spin-1/2 hadron the four quasi GPDs   ${\mathcal H}_{qT}$, 
$\tilde {\mathcal H}_{qT}$, ${\mathcal E}_{qT}$ and $\tilde {\mathcal E}_{qT}$ in the parametrization of ${\mathcal F}_{qT}$ are matched into twist-2 
quark GPDs $H_{qT}$, 
$\tilde H_{qT}$, $ E_{qT}$ and $\tilde  E_{qT}$ with the same perturbative coefficient function $h_{qT}$, respectively.    

\par 
Our results of the gluon contributions in Eq.(\ref{QFAC}) are new. To compare with existing results of quark contributions, we divide the one-loop perturbative functions in Eq.(\ref{QFAC}) into a virtual part 
and a real part. The virtual part is the contribution proportional to $\delta (x-z)$, and the real part is 
the remaining contribution. The latest  results about factorizations of the quark quasi GPDs 
${\mathcal H}_q$, ${\mathcal H}_{qL}$ and ${\mathcal H}_{qT}$ in Eq.(\ref{DQGPD}) are given in Eq.(34-48) of   \cite{MGPD3}     
for the non-singlet case in Eq.(34) . In these results the U.V. divergences are  not subtracted because it is intended to use other renormalization scheme instead
of $\overline{\rm{MS}}$-scheme.  However, a comparison can still be made because that the U.V. divergences are dimensionally regularized. In our results the U.V. divergent contributions can be found in Eq.(\ref{UVQF},30,31). 
They are the same for the three quasi quark GPDs. 
Our virtual- and  real parts of ${\mathcal H}_q$, ${\mathcal H}_{qL}$ and  ${\mathcal H}_{qT}$ are not in agreement with the corresponding results in \cite{MGPD3}, 
respectively.  In the below we discuss the disagreement between real parts in detail. In the discussion we will use our notations of momentum fractions. 

\par 
The convolutions of our real parts in Eq.(\ref{QFAC}) are different than the corresponding convolutions in  \cite{MGPD3}.  It is observed that the real parts do agree with those  corresponding results in \cite{MGPD3}
by taking $x=1$ but disagree by taking $x=-1$. The convolutions are only in agreement in the region of $\xi \le x \le 
1 $ for $\xi>0$.  
It is noted that by taking $x=1$ the real parts are essentially  perturbative contributions of quasi GPDs defined with quark states, in which the initial quark carries the momentum fraction $1+\xi$ and the final quark carries  
the momentum fraction $1-\xi$.  With this observation one of reasons for the disagreement can be the following: 
In \cite{MGPD3} the real parts are obtained from quasi GPDs and GPDs with these mentioned quark states, denoted for instance as $f_1(z,\xi)$. 
For $f_1(z,\xi)$ one always have $\vert \xi \vert <1$ with these  quark states. Therefore, $f_1(z,\xi)$ is in fact unknown or undefined 
for $\vert \xi \vert >1$.  The real parts in \cite{MGPD3} are given by $f_1(z/x,\xi/x)$. 
Then the second variable in $f_1(z/x,\xi/x)$
can be larger than $1$ or smaller than $-1$, i.e., $\vert \xi/x \vert  >1$,  because  $x$ is from $-1$ to $1$ as an integration variable. Therefore, with results of quasi GPDs of the quark states it seems that one can not 
derive the complete results of the real parts. The correct results can be derived with parton states where one 
should take the initial quark 
with the momentum fraction $0< x <1$ instead of $x=1$ and also add the contribution from the initial antiquark for the region  $-1 <x <0$.

\par 
The $\mu$-dependence of quasi quark GPDs is determined by  Eq.(\ref{AZC}). From Eq.(\ref{QFAC}) we can derive 
the $\mu$-dependence of quark GPDs at one-loop level by using results in this work.  For quark GPDs  in the representation or definition used here we have: 
 \begin{eqnarray} 
 \frac{ d F_q (x,\xi,t,\mu) }{ d \ln \mu^2} &=& \frac{\alpha_s}{2\pi}  \int_{-1}^{1} dz \biggr [ 
     {\mathcal P}_{qq} (z,\xi, x) F_q (z,\xi , t,\mu) +{\mathcal P}_{qg} (z,\xi,x) F_g (z,\xi,t, \mu) \biggr ] ,         
 \nonumber\\
\frac{ d F_{qL} (x,\xi,t,\mu) }{ d \ln \mu^2} &=& \frac{\alpha_s}{2\pi}  \int_{-1}^{1} dz \biggr [ 
     {\mathcal P}_{qq} (z,\xi, x) F_{qL} (z,\xi , t,\mu) +{\mathcal P}_{qgL} (z,\xi,x) F_{gL} (z,\xi,t, \mu) \biggr ] ,   
 \nonumber\\
 \frac{ d F_{qT} (x,\xi,t,\mu) }{ d \ln \mu^2} &=& \frac{\alpha_s}{2\pi}  \int_{-1}^{1} dz  
     {\mathcal P}_{qq T} (z,\xi, x) F_{qT} (z,\xi , t,\mu)  ,     
\label{EvolQ}      
 \end{eqnarray} 
 where ${\mathcal P}_{qq}$, ${\mathcal P}_{qg}$,  ${\mathcal P}_{qgL}$,  and ${\mathcal P}_{qqT}$ are evolution kernels. They are given by:
 \begin{eqnarray} 
{\mathcal P}_{qqT} (z,\xi,x) &=& \frac{3 C_F}{2}   \delta (z-x) -\frac{C_F}{2} \biggr [ -\biggr ( \frac{\epsilon (z-x)}{z-x} \biggr )_+ + \frac{\epsilon(z-x)}{z+\xi} -\frac{\vert x+\xi\vert}{(z-x)(z+\xi)} 
\nonumber\\
    && +\delta (x-z) \ln\frac{(z+\xi)^2}{1-z^2} +(\xi \to -\xi) \biggr ],  
\nonumber\\
 {\mathcal P}_{qq} (z,\xi,x) &=& {\mathcal P}_{qqT} (z,\xi,x) + C_F \biggr ( \frac{\vert x+\xi\vert}{4\xi (z+\xi)} -  \frac{\vert x-\xi\vert}{4\xi (z-\xi)}  +\frac{\vert z-x\vert }{2 (z^2-\xi^2) } \biggr ) , 
 \nonumber\\
 {\mathcal P}_{qg} (z,\xi,x) &=& - \frac{1}{2 (z^2-\xi^2)}  
\biggr \{    \epsilon (z-x)   
    \frac{z^2-\xi^2 +2 x^2-2xz}{4 (z^2-\xi^2) }
  +   \frac{\vert x+\xi \vert }{4\xi (z+\xi) } ( z+\xi-2x)     \biggr \}             
    +(\xi \to -\xi), 
 \nonumber\\
  {\mathcal P}_{qgL} (z,\xi,x) &=&  \frac{1}{2 (z^2-\xi^2)}  
\biggr \{    \epsilon (z-x)   
    \frac{z^2+\xi^2 -2xz}{4 (z^2-\xi^2) }
  +   \frac{(z-\xi) \vert x+\xi \vert }{4\xi (z+\xi) }    \biggr \}             
    +(\xi \to -\xi).    
 \end{eqnarray}
 The renormalization of the operator used to define unpolarized quark GPDs has been studied in \cite{BGR,BaBr} in position space. One can transform the evolution kernel into momentum space to obtain the $\mu$-evolution as discussed in \cite{EvoRa}. Our result for the evolution of the unpolarized quark GPDs agrees with that in \cite{BGR,BaBr} after the transformation.   
The renormalization of the operators used to define polarized quark GPDs has been studied in \cite{BGRpol} in position space. 
Our results of the evolutions of polarized quark GPDs also agree with those in \cite{BGRpol} after the transformation except 
a disagreement between overall factors. However, our results agree with those given in \cite{BeRa}. 
From our results, one can correctly re-produce DGLAP equations of quark PDFs, as mentioned later. 
Therefore, the overall factors given here are correct.     
 It is noted that there are different representations or definitions of GPDs in literature. E.g., in \cite{EvoRa}
 the so-called nonforward distributions are introduced, which are GPDs given in the representation other than the one used here. 
 Our evolution equations are given in the symmetric representation of GPDs\cite{MDI}.

\par   
From our results one can derive $P^z$-dependence of quasi quark GPDs. The results are:
 \begin{eqnarray} 
 \frac{\partial {\mathcal F}_q (z,\xi,t)}{\partial \ln (P^z)^2}  &=& \frac{\alpha_s}{2\pi}  \int_{-1}^{1} dx \biggr [ 
      {\mathcal K}_{qq}  (x,\xi,z) F_q (x,\xi , t) + {\mathcal K}_{qg} (x,\xi,z) F_g (x,\xi,t) \biggr ] ,         
 \nonumber\\
 \frac{\partial{\mathcal F}_{qL}  (z,\xi,t)} {\partial \ln (P^z)^2} &=& \frac{\alpha_s}{2\pi}  \int_{-1}^{1} dx \biggr [ 
     {\mathcal K }_{qqL}  (x,\xi,z) F_q (x,\xi , t) + {\mathcal K}_{qgL}  (x,\xi,z) F_{gL}  (x,\xi,t) \biggr ] ,   
 \nonumber\\
 \frac{\partial{\mathcal F}_{qT}^\mu   (z,\xi,t)}{\partial \ln (P^z)^2} &=& \frac{\alpha_s C_F}{2\pi}  \int_{-1}^{1} dx 
     {\mathcal K}_{qqT}  (x,\xi,z) F_{qT}^\mu  (x,\xi , t).   
 \end{eqnarray} 
The coefficient functions ${\mathcal K}$'s are closely related to those in the $\mu$-evolution. At the considered order 
we have: 
 \begin{eqnarray} 
{\mathcal K}_{qq} (x,\xi,z) &=& {\mathcal P}_{qq} (x,\xi,z) -\frac{3 } {2} C_F \delta (x-z) ,  
\quad 
{\mathcal K}_{qqL}  (x,\xi,z) = {\mathcal K}_{qq} (x,\xi,z),          
\nonumber\\
{\mathcal K}_{qqT}  (x,\xi,z) &=& {\mathcal P}_{qqT} (x,\xi,z) -\frac{3 } {2}C_F  \delta (x-z)        , 
\quad 
{\mathcal K}_{qg} (x,\xi, z)      =  {\mathcal P}_{qg} (x,\xi, z),   
\nonumber\\                   
{\mathcal K}_{qgL}  (x,\xi, z)      &=&  {\mathcal P}_{qgL}  (x,\xi, z).                                                    
\end{eqnarray}   
In these $P^z$-dependences power corrections are neglected.

\par 

In the limit of $\Delta^\mu\to 0$, our results about quasi GPDs reduce to those about quasi parton distributions. In the limit, the quark GPDs  $H_q(x,0,0)$, $H_{qL}(x,0,0)$ and $H_{qT}(x,0,0)$ in Eq.(\ref{DGPD})  become corresponding quark parton distributions $q(x)$, $q_L(x)$ and $q_T(x)$, respectively. 
Similarly, the quasi quark GPDs ${\mathcal H}_q(z,0,0)$, ${\mathcal H}_{qL}(z,0,0)$ and ${\mathcal H}_{qT}(z,0,0)$ in Eq.(\ref{DQGPD}) become the quasi quark PDFs $\tilde q(z)$, $\tilde q_L (z)$ and $\tilde q_T (z)$, respectively. 
 In the case of gluon GPDs $H_g(x,0,0)$ and $H_{gL}(x,0,0)$ in Eq.(\ref{GT2}) are  the unpolarized gluon distribution $x g(x)$ and the longitudinal polarized one $x g_L(x)$, respectively. The quasi quark distributions are factorized as:  
 \begin{eqnarray} 
 \tilde q (z)  &=& q (z)  +\frac{\alpha_s}{2\pi}  \int_{-1}^{1} dx \biggr ( 
    C_F c_q (x, z)  q (x ) + c_g (x, z) g (x) \biggr ) ,         
 \nonumber\\
 \tilde q_{L}  (z) &=& q_{ L}  (z) +\frac{\alpha_s}{2\pi}  \int_{-1}^{1} dx \biggr ( 
    C_F c_{qL}  (x,z )   q_{L}  (x) + c_{gL}  (x,z)  g_{L}  (x) \biggr ) ,   
 \nonumber\\
\tilde q_{T}   (z) &=& q_{T}  (x) +\frac{\alpha_s C_F}{2\pi}  \int_{-1}^{1} dx 
     c_{qT}  (x, z)  q_{T}  (x), 
\label{QQFAC}       
 \end{eqnarray} 
 where $c$'s are perturbative coefficient functions. These functions can be obtained by taking $\xi\to 0$ 
 in perturbative coefficient functions of quasi GPDs. They  are:
 \begin{eqnarray} 
c_q (x,z) &=&   -   
   \frac{x-z}{2 x^2}  \biggr ( \epsilon(z)\ln\frac{ \mu^2}{ e (2z P^z)^2 }  + \epsilon(x-z)\ln\frac{  \mu^2}{ e(2(x-z) P^z)^2 }   \biggr ) 
  +\frac{1}{x} \epsilon(z) + c_{qT}(x,z), 
\nonumber\\
c_{qL}  (x, z) &=&c_q (x ,z),          
\nonumber\\
c_{qT}  (x,z) &=&
         \biggr ( \frac{\epsilon(x-z)}{(x-z)}\ln\frac{ (2(x-z) P^z)^2} { e \mu^2}   \biggr)_+  -\frac{\epsilon(x-z)}{x}  \ln \frac{ (2 (x-z) P^z)^2}{e \mu^2}  +\frac{  \vert z  \vert }{x( x-z) }  \ln \frac{ (2z P^z)^2}{ e \mu^2}
\nonumber\\         
  &&               
  +\delta (x-z) \biggr [  - \ln\frac{ (2 P^z)^2} {e \mu^2}\ln\frac{ x^2}{1-x^2}    -2 \ln^2 \vert x \vert  + \ln^2 (1+x) 
 + \ln^2 (1-x)    +\frac{1}{3} \pi^2 \biggr ] ,          
\nonumber\\
c_g(x,z) &=&   \frac{z^2 + (x-z)^2}{4 x^3 } \biggr ( \epsilon(z) 
  \ln\frac{ (2z P^z)^2 }{\mu^2} +\epsilon(x-z)  \ln\frac{ (2(x-z)  P^z)^2 }{\mu^2}   
     \biggr )+ \frac{x^2-2 z^2 }{2x^3} \epsilon(z)
\nonumber\\
    &&   + z\frac{x-z}{x^3} \epsilon(x-z)  ,        
\nonumber\\                   
c_{gL}  (x, z)      &=&    
\frac{x-2z }{4 x^2 } \biggr ( \epsilon (z) \ln\frac{\mu^2}{(2 z P^z)^2}  +  \epsilon(x-z)   
  \ln\frac{\mu^2}{(2 (x-z)  P^z)^2} \biggr )   +  \frac{x-z}{x^2 }  \epsilon(x-z) - \frac{ z}{x^2} \epsilon(z).                                                                                                     
\end{eqnarray}   
Similarly, one can use the above result to derive DGLAP equations of quark PDFs. By noticing the difference 
between the definition of $+$-distributions used here and that commonly used in DGLAP equations, the derived equations agree with quark DGLAP equations.

\par

If we take  $q(x)=\delta (1-x)$ and neglect the gluon contribution in the first equation of Eq.(\ref{QQFAC}), 
we obtain the quasi quark PDF $\tilde q(z)$ of an unpolarized quark at one-loop.  From the current conservation one expects:  
\begin{equation}
   \int_{-\infty}^{\infty} dz \tilde q(z) =1. 
\end{equation} 
But our $\tilde q(z)$ does not satisfy this equation. This is 
because we have already subtracted  the U.V. divergences in the contributions from Fig.2d and 2e before the integration over $z$ here. This is similar to the case of transverse momentum dependent parton distributions, e.g., the unpolarized quark distribution $q(x,k_\perp)$. Formally, one expects that $q(x,k_\perp)$ becomes the parton distribution $q(x)$ after the integration over $k_\perp$: 
\begin{equation} 
  \int d^2 k_\perp q(x, k_\perp) = q(x). 
\end{equation}            
However,  using the result of $q(x,k_\perp)$ of a quark state which can be found in \cite{JMY}, one will find the  integration over $k_\perp$ generates U.V. divergences. With our result of $\tilde q(z)$ the integral also 
contains a U.V. divergence.  The divergence comes from the region with $\vert z\vert \to \infty$.  E.g., for $z>0$  
our $\tilde q(z)$  behaves like $1/z$ for $z\to \infty$. This U.V. divergence can be isolated and subtracted by considering 
the convolution with a test function $t(z)$:
\begin{equation} 
    \int_{-\infty}^{\infty} dz \tilde q(z) t(z),  
\end{equation}       
as suggested in \cite{IJJSZ}.  But, this subtraction and hence the renormalization will depend
on the value of $t(\pm \infty)$.  This will be  against the expectation that the operator used to define $\tilde q(z)$ 
is renormalized by the wave function renormalization constant in the axial gauge as discussed at the end of Sect.3.1.  
 
\par\vskip20pt
\noindent 
{\bf 6. Summary} 
 
We have studied factorizations of quasi quark GPDs with twist-2 parton GPDs, where quasi GPDs in the large hadron momentum are given by convolutions of twist-2 parton GPDs and perturbative coefficient functions. These functions are calculated in this work at one-loop level and are free from collinear- and I.R. 
divergences.  The contributions of twist-2 gluon GPDs are included.  The approach used in this work 
is different than that in existing studies of quasi GPDs. Our results derived with the approach are complete for all quark quasi GPDs and can be used not only 
for hadrons with 1/2-spin but also for hadrons with spins other than 1/2.

\par\vskip40pt

\noindent
{\bf Acknowledgments}
\par
The work is supported by National Natural Science Foundation of P.R. China(No.12075299,11821505, 11847612,11935017 and 12065024)  and by the Strategic Priority Research Program of Chinese Academy of Sciences, Grant No. XDB34000000.

\par

\end{document}